\documentclass[conference]{IEEEtran}
\IEEEoverridecommandlockouts
\usepackage{caption}
\usepackage{subcaption}
\usepackage{cite}
\usepackage{amsmath,amssymb,amsfonts}
\usepackage{graphicx}
\usepackage{textcomp}
\usepackage{xcolor}
\usepackage{multirow}
\usepackage{hyperref}
\usepackage{algorithm}
\usepackage{booktabs}

\usepackage[noend]{algpseudocode}
\usepackage{url}
\def\BibTeX{{\rm B\kern-.05em{\sc i\kern-.025em b}\kern-.08em
    T\kern-.1667em\lower.7ex\hbox{E}\kern-.125emX}}
\begin{document}

\title{Fake News Detection with Heterogeneous Transformer}

\author{\IEEEauthorblockN{Tianle Li\IEEEauthorrefmark{1},
Yushi Sun\IEEEauthorrefmark{2}, Shang-ling Hsu\IEEEauthorrefmark{3},
Yanjia Li\IEEEauthorrefmark{4} and
Raymond Chi-Wing Wong\IEEEauthorrefmark{5}}
\IEEEauthorblockA{Department of Computer Science,
The Hong Kong University of Science and Technology\\
Hong Kong, China\\
Email: \IEEEauthorrefmark{1}tliax@connect.ust.hk,
\IEEEauthorrefmark{2}ysunbp@connect.ust.hk,
\IEEEauthorrefmark{3}shsuaa@connect.ust.hk,
\IEEEauthorrefmark{4}ylifs@connect.ust.hk,
\IEEEauthorrefmark{5}raywong@cse.ust.hk}}


\maketitle
\begin{abstract}
The dissemination of fake news on social networks has drawn public need for effective and efficient fake news detection methods. Generally, fake news on social networks is multi-modal and has various connections with other entities such as users and posts. The heterogeneity in both news content and the relationship with other entities in social networks brings challenges to designing a model that comprehensively captures the local multi-modal semantics of entities in social networks and the global structural representation of the propagation patterns, so as to classify fake news effectively and accurately. 

In this paper, we propose a novel Transformer-based model: HetTransformer to solve the fake news detection problem on social networks, which utilizes the encoder-decoder structure of Transformer to capture the structural information of news propagation patterns. We first capture the local heterogeneous semantics of news, post, and user entities in social networks. Then, we apply Transformer to capture the global structural representation of the propagation patterns in social networks for fake news detection. Experiments on three real-world datasets demonstrate that our model is able to outperform the state-of-the-art baselines in fake news detection.
\footnote{The detailed implementation of our model can be found at: \url{https://github.com/HetTransformer/HetTransformer-model}.}
\end{abstract}
\begin{IEEEkeywords}
Fake News Detection, Transformer, Attention Mechanism, Social Networks, Heterogeneous Graph
\end{IEEEkeywords}
\section{Introduction}
In the era of the rapid growth of social media, fake news on social networks has become a public concern. The dissemination of fake news is rampant due to the convenience of social networks. However, it is difficult for ordinary people to distinguish fake news from real news due to the lack of domain knowledge and limitation of time. Automatic fake news detection methods are necessary for us to effectively and accurately distinguish fake news on social networks.

Many previous methods focus on extracting useful features from news content and then make classification based on these features \cite{yang2012automatic, kwon2013prominent, zhao2015enquiring, wu2015false, castillo2011information, ma2017detect}. Other methods employ deep learning models to capture the necessary features from the news content \cite{qi2019exploiting, schwarz2020emet, udandarao2020cobra, ma2016detecting, ma2018rumor, khattar2019mvae}. A major limitation of these methods is that they ignore the global structural representations in the social network, which have been shown to be useful for fake news detection by previous studies \cite{tang2015line, yuan2019jointly}.

\begin{figure}[t]
   \centering
  \vspace{-0in}
  \includegraphics[width=1.0\columnwidth]{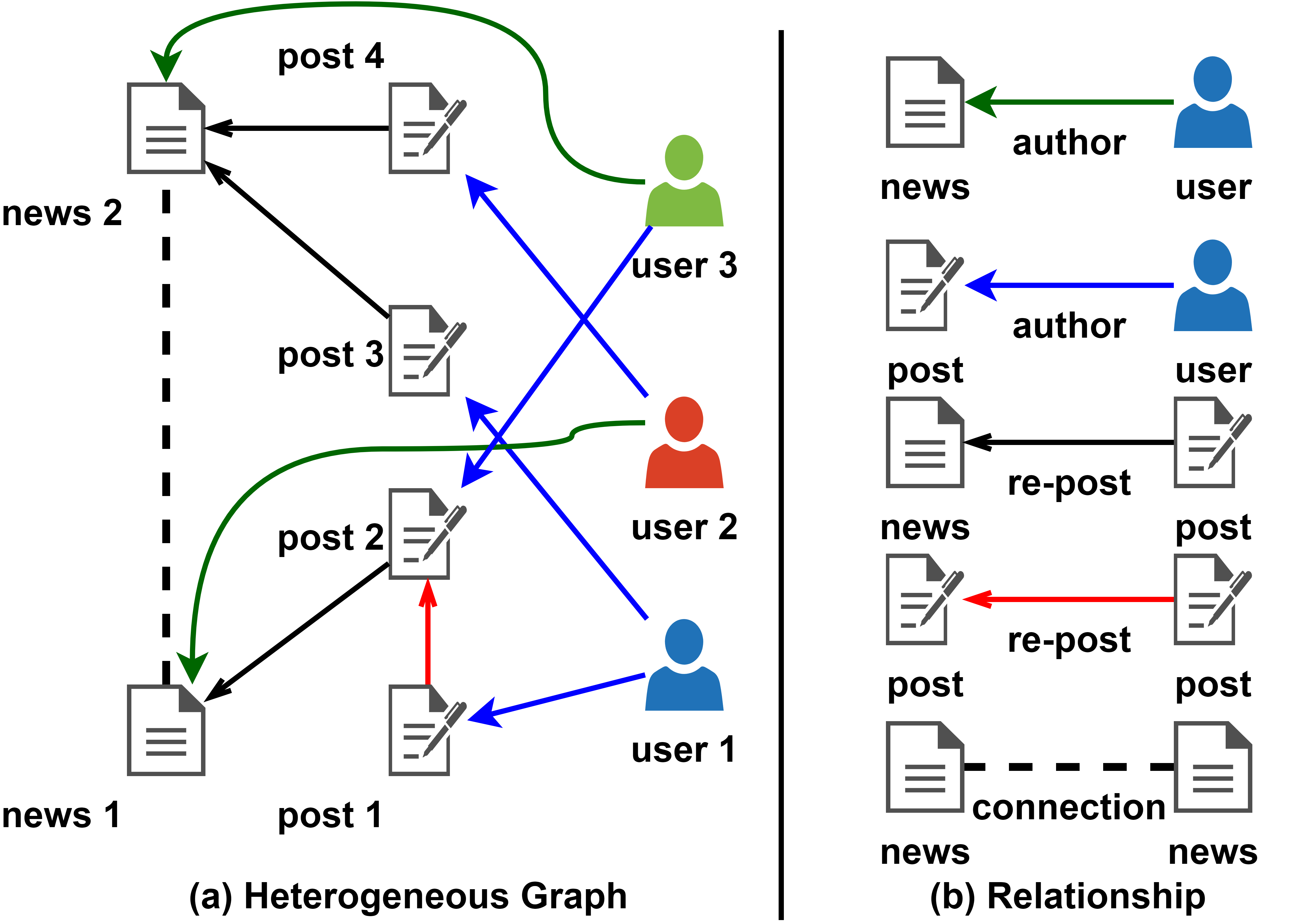}
  \caption{Example of the heterogeneous graph in PHEME. (a) The heterogeneous graph contains three types of nodes with five types of connections. (b) Different types of connections between entities in the heterogeneous graph.}  
~\label{fig:demo}
\end{figure}

The news in social networks does not exist alone. Many other types of entities such as posts and users also exist in social networks and have different types of connections with the news. Each entity contains multi-modal information (e.g., image (in news) and text (in news and post), textual description and user profile statistics (in user)).
The relationship between these entities can be represented as a heterogeneous graph. We provide a brief illustration of the heterogeneous graph for PHEME dataset \cite{derczynski2015pheme} in Fig. \ref{fig:demo}. We can notice that news 1 and news 2 are shared by similar users through the re-post and author relationships. As suggested by \cite{yuan2019jointly}, they are likely to have the same labels (or they are likely to be similar). We use a dash line to denote the \emph{inherent connection} between the two pieces of news. We refer to the news nodes that are connected by this kind of inherent connection as (inherent) news neighbors. This kind of inherent connection between news with a similar neighborhood cannot be captured by the methods that focus on content information only. The rich structural information of the propagation pattern of news embedded in the heterogeneous graph is useful for fake news detection. Therefore, methods that consider both local multi-modal semantics of news and the global structural representation of the propagation patterns of news are suitable for fake news detection tasks in social networks.

Graph Neural Network (GNN) is the most widely used method to process graph structures \cite{gori2005new, kipf2016semi}. However, most of the GNNs are designed for homogeneous graphs, in which different types of nodes are treated as the same \cite{kipf2016semi}, making them not suitable for modeling the structural representation in the heterogeneous social network. Some methods have been proposed to utilize GNNs to deal with heterogeneous graphs \cite{bian2020rumor,  wang2019heterogeneous, zhu2020heterogeneous, yuan2019jointly}. However, they still have major limitations. Firstly, many of them are constrained by graph structures and the meta-path structures. Some of them are constrained by a certain type of graph structures (e.g., the type of graphs that can be decomposed into multiple mini-graphs in \cite{zhu2020heterogeneous}). Some of them require special/additional definitions of the graph structures, which are difficult for people without sufficient domain knowledge to capture (e.g., the additional connections between special user nodes in \cite{yuan2019jointly}). Some of them are constrained to capture the representation based on meta-paths (e.g., the node-level attention defined in \cite{wang2019heterogeneous} ignores the inherent special connections that are not captured by the meta-paths defined in graphs). The adaption of these types of models to other heterogeneous graphs requires domain knowledge. This limitation influences the generalizability of these heterogeneous graph models. Secondly, many of these methods do not fully model the heterogeneity in the heterogeneous graph. Nodes are regarded to be homogeneous in some methods (e.g., \cite{zhu2020heterogeneous}), while edges are regarded to be homogeneous in others (e.g., \cite{bian2020rumor}) making it difficult to fully model the structural representation in the heterogeneous social network. 

Recently, Transformer has drawn researchers' attention to deal with heterogeneous graphs. Hu et al. proposed Heterogeneous Graph Transformer (HGT) to utilize the encoder structure of Transformer to capture the structural representation of heterogeneous graphs \cite{hu2020heterogeneous}. However, this method is not appropriate for solving the problems in fake news detection: On one hand, the multi-modal semantics in social networks cannot be captured using the textual feature encoders in HGT. On the other hand, the encoder-only Transformer structure used by HGT only captures the neighborhood embedding of the target node which can be improved to an encoder-decoder structure to better extract the global structural representation in social networks. 

In light of the point that news nodes sharing the similar neighborhood are likely to have the same labels \cite{yuan2019jointly}, we propose HetTransformer model. The encoder-decoder Transformer structure of HetTransformer can better capture the structural representation of the propagation pattern in the sense that the encoder part captures the representation of the target news node and its neighborhood which consists of (inherent) news neighboring nodes, post nodes, and user nodes, while the decoder structure decodes the representation captured by encoder to the target news node and its (inherent) news neighbors. Unlike HGT, feature encoders are designed in our model to capture the local multi-modal semantics of each entity in social networks. The multi-modal features captured are passed to the Transformer in order to gain the global structural representation of news propagation patterns and then used for fake news classification. Our major contributions can be summarized as follows:
\begin{itemize}
    \item We leverage Transformer structure to aggregate structural representation in social networks. To the best of our knowledge, we are the first to study employing Transformer structure in social networks for fake news detection.
    \item We propose a novel idea to use an encoder structure of Transformer to encode the content of the target news and its neighborhood and a decoder structure to decode the representation gained from encoder to the target news node and its (inherent) news neighbors.
    \item We propose HetTransformer model that not only captures the local multi-modal semantics of each entity in social network but also deals with the global structural representation in the heterogeneous social network.
    \item We conduct a series of experiments on three real-world datasets. The experimental results demonstrate that our proposed HetTransformer achieves superior performance over the state-of-the-art baselines on the fake news detection task.
\end{itemize}

The rest of this paper will be organized as follows. In Section \ref{sec: related work}, we introduce the related work. In Section \ref{sec: problem definition}, we formally introduce the definition of the heterogeneous graph and the fake news detection problem. We provide detailed introduction of our proposed model in Section \ref{sec: methodology}. In Section \ref{sec: experiment}, we present the experimental results on three real-world datasets. Finally, we conclude our work in Section \ref{sec: conclusion}.
\section{Related Work}
\label{sec: related work}
In this section, we briefly introduce the following related types of fake news detection methods: Traditional methods, Graph Neural Network methods, and Transformer-based methods.
\subsection{Traditional methods}
Traditional methods utilize features extracted from the news content and/or user profiles to train a classifier for fake news detection.

Some early studies utilize hand-crafted features extracted from news (e.g., \cite{yang2012automatic, kwon2013prominent, zhao2015enquiring, wu2015false}). A decision tree classifier was designed by \cite{castillo2011information} to classify the news type based on the hand-crafted linguistic features. SVM with RBF kernel was introduced by \cite{yang2012automatic} to detect fake news based on the overall statistics of news and users without using the temporal information. Similarities between the propagation relationships of different news posts are utilized in \cite{ma2017detect} to examine the credibility of news. These traditional methods largely rely on hand-crafted features extracted from news content, user profiles, and propagation relationship, which are dataset-specific and require large human efforts and domain knowledge to design specific feature extraction rules for every single dataset, inducing low generalizability.

In order to resolve the problem of hand-crafted methods, some existing deep learning models were adapted to detect fake news based on news content (e.g. \cite{qi2019exploiting, schwarz2020emet, udandarao2020cobra}). Long Short Term Memory (LSTM) and Gated Recurrent Unit (GRU) were employed by \cite{ma2016detecting}. The usage of Recursive Neural Networks (RvNN) in fake news detection was explored by \cite{ma2018rumor}. Visual features of the news content were captured using Convolutional Neural Network (CNN) in \cite{khattar2019mvae}. Although these methods captured the textual information and visual information from news content and combined the information from multi-modality to detect fake news, they ignored the propagation relationship between news and users, and thus failed to fully capture the social context structural representation in social networks, causing disadvantages in detecting fake news.

\subsection{Graph Neural Network methods}
In addition to the news content, the propagation structure and social context of the posted news also play a vital role in detecting fake news on social media. Graph-oriented approaches were commonly employed in previous studies to encode the social structure information (e.g. \cite{lu2020gcan, zhou2020graph, huang2019deep, wen2020asa, rao2021suspicious, yang2020rumor}). Compared to the previous methods, GCN \cite{kipf2016semi} is capable of aggregating context features in social networks by adopting spectral-domain convolution. Bian et al. proposed a Bi-Directional Graph Convolutional Networks (Bi-GCN) to reveal the characteristics of rumors from both top-down and bottom-up propagation \cite{bian2020rumor}. Although the direct adoption of GCN is a tempting solution, previous studies \cite{xu2018powerful, zhu2020heterogeneous, fan2020metagraph} have shown that GCN is inferior in fake news detection tasks due to its incapability of distinguishing different types of relations. The inherent homogeneity of GCN induces challenges for GCN-based models to fully capture the heterogeneity of data associations in social networks. 

To address the challenge, solving news classification problems by modeling the network structure as heterogeneous graphs becomes a popular approach in fake news detection \cite{ren2020adversarial}, fraud detection \cite{wen2020asa, zhu2020heterogeneous}, illicit traded product identification \cite{fan2020metagraph}, and suspicious massive registration detection \cite{rao2021suspicious}. 
Recent approaches proposed different types of Heterogeneous Graph Neural Networks (HetGNNs), processing heterogeneous types of nodes or relations separately:
GLAN \cite{yuan2019jointly} and HMGNN \cite{zhu2020heterogeneous} conduct rumor detection with different variants of hierarchical, attention-based HetGNNs to capture graph structural information. The local semantics and the global holistic graph structures are jointly encoded in GLAN. However, GLAN only focused on the structural information of the propagation pattern of news, without aggregating multi-modal information of news in social networks. HMGNN focused on capturing heterogeneous semantics from edges instead of nodes in the social network graph. As a result, the HMGNN can not extract the heterogeneity of node content, and thus could not fully capture the heterogeneity semantics in social networks.
\subsection{Transformer-based methods}
Recently, models that use Transformer have been proposed. Jwa et al. proposed exBAKE to directly apply BERT to process textual content of news and then finetune the model to perform fake news detection \cite{jwa2019exbake}. Despite the superior capability of extracting representation of news text, exBAKE ignored the propagation pattern of news, and thus could not capture the social context information of news, which is harmful for fake news detection. Hu et al. proposed Heterogeneous Graph Transformer (HGT) to use Transformer component to capture the graph representation and then perform node classification and link prediction tasks \cite{hu2020heterogeneous}. However, HGT focused on capturing graph-level heterogeneity (i.e., the different types of nodes) and the structural information of the graph instead of the node-level heterogeneity (i.e., the multi-modal contents of nodes). Due to the multi-modal property of social networks, a more fine-grained model that can capture graph-level heterogeneity, structural information and node-level heterogeneity is required in the fake news detection task. Besides, the Transformer structure in HGT only considers to encode the neighborhood of the target node, without employing a special decoder to refine the representation with the target node and its (inherent) neighbors.

In this paper, our model utilizes heterogeneous content encoders to capture the local multi-modal semantics from news, post, and user entities. Then, we apply an encoder-decoder structure of Transformer to aggregate the global structural representation of social networks so as to classify fake news.
\section{Problem Definition}
\label{sec: problem definition}
\subsection{Terminology Definition}\label{sec:terminology}
The datasets used in our work contain news articles, posts, and social media users, which can be represented as three different types of nodes in our heterogeneous graphic network and can be linked with edges based on the connections among them. We define our heterogeneous graph formally as follows: 

\textit{DEFINITION 1:}
The heterogeneous graph can be defined as $\mathcal{G} = (\mathcal{V}, \mathcal{E})$, where the node set $\mathcal{V}=\mathcal{N}\cup\mathcal{P}\cup\mathcal{U}$. Here, $\mathcal{N}, \mathcal{P}, \mathcal{U}$ represent news articles, posts, and social media users, respectively. The edge set $\mathcal{E}=\mathcal{E}_{n,p}\cup\mathcal{E}_{n,u}\cup\mathcal{E}_{p,u}\cup\mathcal{E}_{p, p}\cup\mathcal{E}_{u, u}$ contains the edges ``between news articles and posts'' and ``between posts and posts'' based on ``re-post'' relationship (denoted by $\mathcal{E}_{n,p}$ and $\mathcal{E}_{n, u}$), the edges ``between news and users'' and ``between posts and users'' based on ``author'' relationship (denoted by $\mathcal{E}_{n,u}$ and $\mathcal{E}_{p,u}$ respectively), and the edges ``between users and users'' based on ``follow'' relationship (denoted by $\mathcal{E}_{u,u}$). \footnote{The relationships in \textit{DEFINITION 1} are based on PolitiFact and GossipCop datasets. The relationships defined in PHEME dataset consist of the ``re-post'' between post - news and between post - post; and the ``author'' between post - user and news - user only without ``follow'' relationship.}

\textit{DEFINITION 2:}
(News articles): The news article set is represented as $\mathcal{N} = \{n_1, n_2, \ldots, n_m\}$. For each news $n_i\in\mathcal{N}$, it contains the textual and visual contents and a credibility label $y_i\in\{0, 1\}$, where $y_i=0$ stands for fake news and $y_i=1$ stands for real news.

\textit{DEFINITION 3:}
(Posts): The post set is represented as $\mathcal{P}=\{p_1, p_2, \ldots, p_k\}$. For each post $p_i\in\mathcal{P}$, it contains the textual contents. 
Post features (e.g., number of forward and number of comments) are also involved for some datasets.

\textit{DEFINITION 4:}
(Social Media Users): The user set is presented as $\mathcal{U}=\{u_1, u_2, \ldots, u_n\}$. For each user $u_i\in\mathcal{U}$, it contains the user profile information, which includes user description, number of followers, geographic location, verification status, etc.
\subsection{Problem Formulation}\label{sec:problem formulation}
We define fake news as the news articles that contain misleading information to the public, as is widely adopted in recent studies \cite{fakenewsdefinition}. This definition can be stated formally as follows:

\textit{DEFINITION 5: }(Fake News): In social media platforms, a piece of fake news is a news article that satisfies one of the three conditions: (1) contains statements that contradict with the fact. (2) contains images that are irrelevant to the news content. (3) contains images that are intentionally manipulated.

The problem addressed in this paper is how to utilize the content and the social structure information to identify a news article as fake or real, which can be considered as a binary classification problem. The formal definition of our problem is given as follows:

\textit{PROBLEM 1: }Given a heterogeneous graph $\mathcal{G}=(\mathcal{V}, \mathcal{E})$ with a set of news article nodes $\mathcal{N}\subset\mathcal{V}$, learn a classifier function: $\mathcal{N}\xrightarrow{}\mathcal{Y}$ to correctly classify each news node $n_i\in\mathcal{N}$ to fake news ($y_i=0$) or real news ($y_i=1$).  
\section{Methodology}
\label{sec: methodology}
\begin{figure*}[t]
   \centering
  \vspace{-0.6in}
  \includegraphics[width=1.0\textwidth]{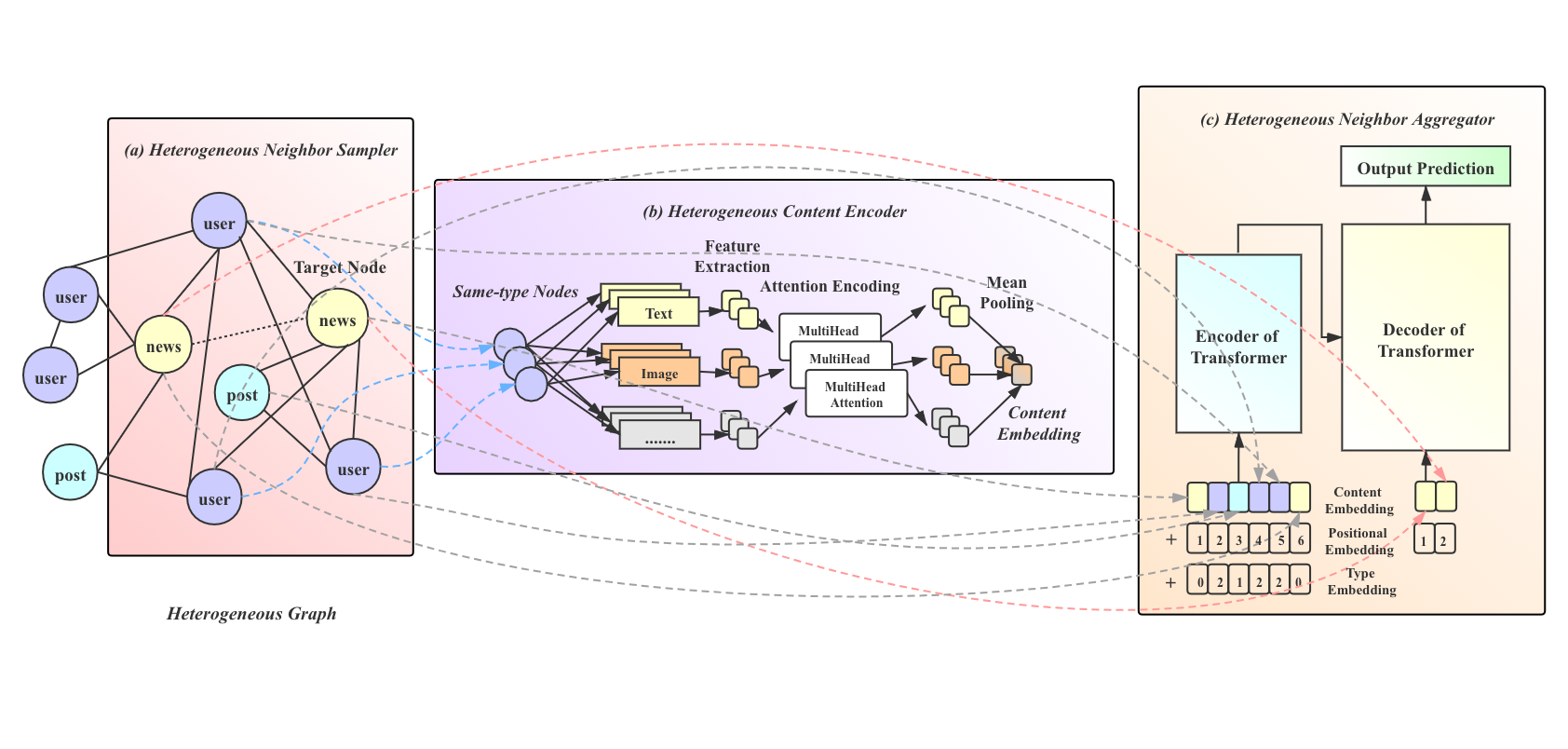}
  \vspace{-0.8in}
  \caption{HetTransformer model architecture: (a) Sample the neighbors of a target news node; (b) Aggregate the heterogeneous content among the same-typed neighbors with MultiHeadAttention; (c) Aggregate heterogeneous-type neighbor representations obtained in (b) for the targeted node and make prediction.}
~\label{fig:HG} \label{fig:architecture}
\end{figure*}

In this section, 
we first give an overview of HetTransformer model (Section~\ref{subsec:overview}) and then introduce the details of the model architecture, including three major components: (1) the heterogeneous neighbor sampler (Section~\ref{sec:rwr}), (2) the heterogeneous content encoder (Section~\ref{subsec:ContentEncoder}), (3) 
the heterogeneous neighbor aggregator (Section~\ref{sec: neighbors aggregator}). 
%
Fig. \ref{fig:architecture} shows the overall procedure.
Then, we describe our model training in Section~\ref{subsec:training}
and the model analysis in Section~\ref{subsec:analysis}.

\subsection{Model Overview} \label{subsec:overview}
The input data to our model is a heterogeneous graph $\mathcal{G}$ which contains news nodes, post nodes, and user nodes. Each news node on the graph takes turns to be the root node of random walk for multi-type neighbor sampling. Then, the heterogeneous content encoder produces an embedding vector to represent each node by encoding their text, images, and other features. The embeddings of the target news node and its heterogeneous-type neighbors are fed into the encoder of the Transformer with the order in random walk sampling. The embeddings of the target news nodes, along with its news-type neighbors, and the output from the encoder of the Transformer are fed into the decoder of the Transformer. Finally, the output of the Transformer decoder is used to train the fake news classifier to predict the label \textit{y} of each news node. 

\subsection{Heterogeneous Neighbor Sampler} \label{sec:rwr}
Following DeepWalk \cite{perozzi2014deepwalk} and HetGNN \cite{zhang2019heterogeneous}, we sample heterogeneous neighbors based on random walk with restart (RWR) \cite{tong2006fast}. Each news node takes turns to be the root node $n_i$, and for each root node, we repeat the following procedure for $\mathcal{T}$ times:
\begin{enumerate}
    \item With probability $p$, jump back to the root node.
    \item Select a first-order neighbor of the current node uniformly at random.
    \item Record the selected node as a neighbor of the root node.
    \item Walk to the selected node.
\end{enumerate}

\newcommand{\Rand}{\mathrm{Rand}}
\newcommand{\Sample}{\mathrm{Sample}}
\newcommand{\SortMostFrequent}{\mathrm{SortMostFrequent}}
A formal description in pseudo code can be found in Line 1 to 10 of Algorithm \ref{algo:hetT}, where ${\Rand}(0,1)$ selects a random number from $U(0,1)$, the standard uniform distribution between 0 and 1, $\Sample(N(v))$ draws one neighbor from $N(v)$, the set of neighbors of $v \in \mathcal{V}$.
After $\mathcal{T}$ iterations, the top $\gamma$ most frequent neighbors are kept and sorted by frequency for each news node, which is denoted as $\SortMostFrequent(\cdot, \gamma)$.

The news neighbors, post neighbors, and user neighbors of the root node generated through RWR process can be obtained even if some of them are not directly connected to the root node.

\subsection{Heterogeneous Content Encoder}
\label{subsec:ContentEncoder}

\subsubsection{Content Encoding}\label{sec: content encoder}
The information from different modals are encoded with different methods. For text and image content, state-of-the-art pre-trained models are employed. As for the other features of the nodes, we follow the approaches described in \cite{lu2020gcan}.

\paragraph{Text Encoding} We fine-tune RoBERTa model \cite{liu2019roberta} to encode all textual content in the datasets and further encode the news content with a T5 \cite{raffel2020exploring} model.

\paragraph{Image Encoding} We utilize ResNet \cite{resnet} to encode image content.

\paragraph{Other Feature Encoding} We define feature vector $X_i = [x_1, x_2, \ldots, x_n]\in\mathbb{R}^{n}$ to represent the discrete and numerical features in user characteristics.

\newcommand{\MultiHead}{\mathrm{MultiHead}}
\newcommand{\HetT}{\mathrm{HetT}}
\newcommand{\neigh}{\mathit{neigh}}
\newcommand{\dec}{\mathit{dec}}
\newcommand{\enc}{\mathit{enc}}
\newcommand{\epochs}{\mathit{epochs}}
\newcommand{\Transformer}{\mathrm{Transformer}}

\begin{algorithm}[tb!]
\caption{Heterogeneous Transformer (HetTransformer)}\label{algo:hetT}
\textbf{Input}: graph $\mathcal{G}=(\mathcal{V}, \mathcal{E})$, labeled news nodes $\mathcal{N}$ with labels $\mathcal{Y_N}$, restart rate $p$, walks per node $\mathcal{T}$, number of neighbors $\gamma$, number of epochs $\epochs$ \\
\textbf{Output}: predicted labels of news nodes $Y_N=\HetT(\mathcal{N})$, \space $ \mathcal{N} \subset V$
\begin{algorithmic}[1]
\For{$n_i \in \mathcal{N}$}
    \State $N'(n_i) \gets \emptyset$
    \State $v \gets n_i$
    \For {$t \gets 1, 2, ..., \mathcal{T}$}
        \If {$\Rand(0, 1) < p$}
            \State $v \gets n_i$
        \EndIf
        \State $w \gets \Sample(N(v))$
        \State $N'(n_i) \gets N'(n_i) \cup \{w\}$
        \State $v \gets w$
    \EndFor
    \State $N''(n_i) \gets \SortMostFrequent(N'(n_i), \gamma)$
\EndFor. 
\For{$k \gets 1, 2, \ldots, \epochs$}
    \State Project heterogeneous attribute features of $v \in \mathcal{V}$ to the same dimensional space.
    \For{$n_i \in \mathcal{N}$}
        \State Concatenate the same-type attribute features within the same-type nodes $\{n_i\}; \neigh_{n}, \neigh_{p}, \neigh_{u} \subset N''(n_i)$.
        \State $E_{n_i} \gets \MultiHead^{(n)}(n_i)$
        \State $E^{(i)}_{\neigh_{n}} \gets \MultiHead^{(n)}(\neigh_{n})$
        \State $E^{(i)}_{\neigh_{p}} \gets \MultiHead^{(p)}(\neigh_{p})$
        \State $E^{(i)}_{\neigh_{u}} \gets \MultiHead^{(u)}(\neigh_{u})$
        \State Concatenate $E_{n_i}$, $E^{(i)}_{\neigh_{n}}$, $E^{(i)}_{\neigh_{p}}$, $E^{(i)}_{\neigh_{u}}$ with frequency order in $N''(n_i)$ and obtain $H^{(i)}_{\enc}$;
        \State Concatenate $E_{n_i}$, $E^{(i)}_{\neigh_{n}}$ and obtain $H^{(i)}_{\dec}$;
        \State Predict $\hat{y_i} = \Transformer(H^{(i)}_{\enc}, H^{(i)}_{\dec})$ 
    \EndFor
\EndFor
\end{algorithmic}
\end{algorithm}

\subsubsection{Heterogeneous Content Aggregator}\label{sec:Content Aggregation}
After extracting the heterogeneous features in a node as Section \ref{sec: content encoder} and generating the neighbors with RWR in Section \ref{sec:rwr}, one challenge is to fuse the heterogeneous modal information and extract an informative node representation for heterogeneous-type nodes respectively. In order to combine the features of different attributes, we concatenate the same-type attribute features (e.g., textual content, image) from the same-type neighbor nodes $neigh_{n}, neigh_{p}, neigh_{u} \subset N''(n)$ \footnote{We deal with the target news node $n$ and its news-type neighbors $neigh_{n}$ separately.} generated by RWR respectively and apply a type-specific transformation matrix $M^{\phi_i}_k \in \mathbb{R}^{d \times d_{ik}}$, where $\phi_i$ represents an attribute type ${i}$; $k \in \{n, p, u\}$ indicates the type of the node, $d$ denotes the unified dimension size projected to, and $d_{ik}$ represents the embedding size of attribute type $i$ from node type $k$. The projection process can be formulated as follows:

\newcommand{\Concat}{\mathrm{Concat}}
\newcommand{\emb}{\mathit{emb}}
\[Q^{{\phi}_{i}}_{k} = \Concat({\emb^{{\phi}_{i}}_{k,1}}, ..., {\emb^{{\phi}_{i}}_{k,{m}_k}})\ \tag{1} \label{eq:feature_encoder}\]
\[{Q^{{\phi}_{i}}_{k}}' = M^{{\phi}_i}_{k}  Q^{{\phi}_{i}}_{k} \tag{2} \label{eq:attri_project}\]
where $emb^{{\phi}_{i}}_{k,j} \in \mathbb{R}^{d_{ik}}$ is the embedding feature of $\phi_i$ from node $n_j \in neigh_{k}$, ${m}_k$ indicates the number of nodes in $neigh_{k}$, $Q^{{\phi}_{i}}_{k} \in \mathbb{R}^{{m}_k \times d_{ik}}$ and ${Q^{{\phi}_{i}}_{k}}' \in \mathbb{R}^{{m}_k \times d}$ are the concatenated features of $\phi_i$ type attribute in nodes of type $k$ before and after the unified projection operation. Then, we apply multi-head self-attention mechanism \cite{vaswani2017attention} to different modal features with the same-type neighbor nodes respectively:

\[E^{{\phi}_{i}}_{k} = \MultiHead({Q^{{\phi}_{i}}_{k}}') \tag{3} \label{eq:feature_encoder}\]
where $E^{{\phi}_{i}}_{k} \in \mathbb{R}^{{m}_k \times d}$ is the encoded attribute features of $\phi_i$ with weighted importance from $k$ type of neighbors after the multi-head attention layers, so that the learned representation of each attribute feature for a single node is aware of its social context in $N''(n_i)$ generated by RWR. The multi-head attention mechanism is formulated as:

\newcommand{\Attention}{\mathrm{Attention}}
\newcommand{\SoftMax}{\mathrm{Softmax}}
\newcommand{\head}{\mathit{head}}
\begin{gather}
 \label{eq:multihead}
  \MultiHead(Q, K, V) = \Concat(\head_1, ..., \head_h)W^O \notag\\
  head_{i} = \Attention(Q W^Q_i, K W^K_i, V W^V_{i}) \tag{4}\\
  \Attention(Q, K, V) = \SoftMax(\frac{QK^T}{\sqrt{d_k}})V \notag
\end{gather}


To combine the multi-modal features within a single node, a mean pooling layer is employed after the multiple multi-head attention layers:
\newcommand{\MeanPool}{\mathrm{MeanPool}}
\[E_{\neigh_{k}} = \MeanPool(E^{{\phi}_{1}}_{k}, ..., E^{{\phi}_{q_k}}_{k}) \tag{5} \label{eq:meanpool} \]
where $q_k$ is the number of different attributes in nodes of type $k$; $E_{\neigh_{k}} \in \mathbb{R}^{{m}_k \times d}$ is the aggregated node representations for nodes of type $k$. For each $v_k \in \neigh_k$ we can obtain its node embedding from $E_{\neigh_{k}}$ by indexing its local RWR order in $\neigh_{k}$. Similarly, we can deduce $E_{n} \in \mathbb{R}^{d}$, which is the node representation of the target news node with the same procedure.

\subsection{Heterogeneous Neighbor Aggregator}\label{sec: neighbors aggregator}
After we obtain the content embeddings of the heterogeneous neighbors $E^{(i)}_{\neigh_{n}}$, $E^{(i)}_{\neigh_{p}}$, $E^{(i)}_{\neigh_{u}}$, and $E_{n_i}$ for the target news node $n_i$ itself,  we leverage Transformer to aggregate the heterogeneous neighbors. Instead of dealing with the heterogeneous-typed neighbor nodes separately, we process them simultaneously to maintain the original news propagation pattern extracted from RWR. 
We concatenate the content embeddings of the target news $E_{n_i}$ along with each of the heterogeneous neighbor nodes with the order from $N''(n_i)$ determined by RWR-based strategy as $H^{(i)}_{enc}$ in Equation \eqref{eq:concatenate-enc}, where $l=m_n+m_p+m_u$ indicates the total number of neighbor nodes in $N''(n_i)$,
\[
H^{(i)}_{enc} = \Concat(E_{n_i}, {E^{(i)}_{\neigh_{k}}}_1, ..., {E^{(i)}_{\neigh_{k}}}_l) \tag{6} \label{eq:concatenate-enc}
\]
\[
H^{(i)}_{dec} = \Concat(E_{n_i}, {E^{(i)}_{\neigh_{n}}}_{1}, ..., {E^{(i)}_{\neigh_{n}}}_{m_n}) \tag{7} \label{eq:concatenate-dec}
\]
where $H^{(i)}_{enc} \in \mathbb{R}^{(l+1) \times d}$. We feed the processed heterogeneous neighbor embedding sequence $H^{(i)}_{enc}$ into the encoder of Transformer together with their corresponding positional embeddings and type embeddings to indicate the realistic propagation sequence in the social network $\mathcal{G}$. In the meantime, the content embeddings of the target news $N_{n_i}$ and its ordered news neighbors content embeddings $H^{(i)}_{\neigh_{n}}$ are concatenated as $H^{(i)}_{dec}$ in equation \eqref{eq:concatenate-dec}, where $H^{(i)}_{dec} \in \mathbb{R}^{(m_n+1) \times d}$, and feed into the decoder of the Transformer in sequence.


\subsection{Model Training}
\label{subsec:training}
\label{sec: predictor}
The first feature embedding $Rep_{n_i}$ from the output of the decoder of Transformer is treated as the aggregated representation of the target news node ${n_i}$. We pass ${Rep_{n_i}}$ to a fully connected layer with activation function and finally make the prediction as 
\newcommand{\Sigmoid}{\mathrm{Sigmoid}}
\newcommand{\ReLU}{\mathrm{ReLU}}
\newcommand{\Rep}{\mathit{Rep}}
\newcommand{\out}{\mathit{out}}
\[
\hat{y_i} = \Sigmoid(\ReLU(\Rep_{n_i}W_{\out} + b_{\out})) \tag{8} \label{eq:output}
\]
where $W_{out}$ is the weight matrix with learnable parameters, and $b_{out}$ is the output bias. The loss function is utilized to minimize the cross-entropy value:
\[
\mathcal{J} = -{(y\log(\hat{y}) + (1 - y)\log(1 - \hat{y}))} \tag{9} \label{eq:loss}
\]

\subsection{Model Analysis}
\label{subsec:analysis}
We analyze the time complexity of the proposed model in this section.
Consider a graph $\mathcal{G}=(\mathcal{E}, \mathcal{V})$ with a subset of nodes for classification $\mathcal{N} \subset \mathcal{V}$, which is the set of news nodes $|\mathcal{N}| \ll |\mathcal{V}|$ in our case, and each node has $I$ types of heterogeneous input. 
Suppose there are $\mathcal{T}$ RWR iterations, $H$ self-attention heads, $L$ layers of Transformer encoders and decoders, and $\delta$ maximum dimension for all representations and input features.
Heterogeneous neighbor sampling takes $O(\mathcal{T}|\mathcal{N}|)$ time, and  $\gamma$ sampled neighbors are kept and sorted for each news node, which takes $O(|\mathcal{N}|\gamma \log(\gamma))$ time.
Since this is a one-off pre-processing step, it does not need to be repeated in the training loop.
Now, we consider the steps and their time complexity in one iteration:
It takes $O(|\mathcal{V}| I \delta^2)$ to project all types of attribute embeddings of all nodes. 
Aggregating heterogeneous content with multi-head self-attention takes $O(H |\mathcal{V}|(I^2 \delta + I \delta^2))$. 
Aggregating heterogeneous neighbors with Transformer takes $O(H L |\mathcal{N}| (\gamma^2 \delta + \gamma \delta^2))$ time. 
Overall, the time complexity of HetTransformer for one iteration is $O(|\mathcal{V}| I \delta^2 + H |\mathcal{V}|(I^2 \delta + I \delta^2) + H L |\mathcal{N}| (\gamma^2 \delta + \gamma \delta^2)) = O(H (|\mathcal{V}|(I^2 \delta + I \delta^2) + L |\mathcal{N}| (\gamma^2 \delta + \gamma \delta^2)))$.
More generally, if HetTransformer is used for other node classification tasks on some graph with heterogeneous nodes and heterogeneous node content, we may assume $\mathcal{N}=\mathcal{V}$ and simplify the complexity as $O(H|\mathcal{V}|(I^2 \delta + I \delta^2 + L \gamma^2 \delta + L \gamma \delta^2))$.
If $H$, $I$, and $L$ are small constants, we may further simplify the complexity to be $O(|\mathcal{V}|(\gamma^2 \delta + \gamma \delta^2))$.
Together with the time needed for RWR, it takes $O(|\mathcal{V}|(\mathcal{T} + \gamma \log(\gamma) + \gamma^2 \delta + \gamma \delta^2))$ in total.
If HetTransformer is used on some graph on its original 1-hop neighbors instead of sampling $\gamma$ neighbors, which does not require RWR, its time complexity is $O(|\mathcal{E}|(|\mathcal{V}|\delta + \delta^2))$.

\section{Experiments}\label{sec:experiments}
\label{sec: experiment}
\subsection{Datasets}\label{sec:dataset}
To evaluate the performance of HetTransformer, we select three benchmark datasets, namely PolitiFact and GossipCop from FakeNewsNet \cite{shu2020fakenewsnet} \footnote{FakeNewsNet datasets are at \url{https://github.com/KaiDMML/FakeNewsNet}.} and 
PHEME \cite{derczynski2015pheme} \footnote{PHEME is available at \url{https://doi.org/10.6084/m9.figshare.4010619.v1}.} for the experiments. The news entities that have empty content are excluded from the datasets.
Table \ref{tab:dataset} shows the statistics of the datasets after excluding the empty news entities.\footnote{Twitter Developer Policy \cite{twitter_privacy_policy} forces developers to obtain their own copies of datasets, so the dataset statistics may differ by publications.}
PolitiFact and GossipCop, the first and the second dataset, focus on widely-spread political and celebrity news, respectively, in English-speaking communities, especially in the U.S. society.
The third dataset, PHEME \cite{derczynski2015pheme}, comprises of facts and rumors associated with different breaking news for rumor analyses in journalism. 

\begin{table}[tb!]
\caption{Statistics of the datasets$^{\mathrm{a}}$}
\begin{center}
\begin{tabular}{l r r r}
\hline
Statistic & PolitiFact & GossipCop & PHEME \\
\hline
\# Fake News    & 385 &  4,913 & 2,402  \\
\# Real News    & 401 & 15,446 & 4,023 \\
\# Total News   & 786 & 20,359 & 6,425 \\
\# Posts        & 440,467 & 1,192,766 & 98,929 \\
\# Users        & 468,210 & 429,628 & 51,043 \\
\hline
\multicolumn{3}{l}{$^{\mathrm{a}}$``\#'' denotes ``the number of.''}
\end{tabular}
\label{tab:dataset}
\end{center}
\end{table}

\subsection{Baselines}\label{sec:baselines}
We select several state-of-the-art models to be our baselines and compare them with our proposed model. Apart from that, we also implement two off-the-shelf supervised learning methods \cite{yang2012automatic, castillo2011information} for more comprehensive comparison. The baselines can be categorized into three categories: Hand-crafted feature-based methods, GNN-based methods, and Transformer-based methods.\\
\textit{Hand-crafted feature-based methods}
\begin{itemize}
    \item \textbf{SVM-RBF} \cite{yang2012automatic}: An SVM classifier with RBF kernel function, utilizing hand-engineered features. 
    \item \textbf{DTC} \cite{castillo2011information}: A J48 Decision Tree classifier, using handcrafted features.\end{itemize} 
    \textit{GNN-based methods}
    \begin{itemize}
    \item \textbf{IARNet} \cite{yu2020iarnet}: A heterogeneous GNN with type-specific attention-based graph convolution for fake news detection.
    \item \textbf{HMGNN} \cite{zhu2020heterogeneous}: An attention-based heterogeneous GNN that constructs hyper-nodes to share information among similar mini-graphs for fraud detection.
    \item \textbf{HetGNN} \cite{zhang2019heterogeneous}: A heterogeneous GNN with hierarchical, type-specific, BiLSTM-based graph convolution.
    \end{itemize}
    \textit{Transformer-based methods}
    \begin{itemize}
     \item \textbf{BERT} \cite{devlin2019bert}: A Transformer-based language model that produces the state-of-the-art performance for a range of tasks.
    \item \textbf{HGT} \cite{hu2020heterogeneous}: A Transformer-based model with type-specific message passing and aggregation. Encoder structure of Transformer is utilized to capture the graph structural representations.
\end{itemize}

Among the baselines for heterogeneous graphs, HMGNN \cite{zhu2020heterogeneous} is designed for heterogeneous edges \textit{but} homogeneous nodes, while HetGNN \cite{zhang2019heterogeneous}, IARNet \cite{yu2020iarnet}, and HGT \cite{hu2020heterogeneous} are designed for graphs with heterogeneous edges and nodes.
Besides, BERT \cite{devlin2019bert} and HGT \cite{hu2020heterogeneous} utilize Transformer: The former uses it as informative pre-trained text embeddings, while the latter makes use of it for graph convolution.

\subsection{Evaluation Metrics}
We use Accuracy, Precision, Recall, and F1 score as the evaluation metrics. 
\begin{itemize}
    \item Accuracy is defined as the ratio between the number of correctly predicted news and the total number of news.
    \item Precision ($P$) is defined as the ratio between the number of true positives and the number of positive elements.
    \item Recall ($R$) is defined as the ratio between the number of true positives and the number of relevant elements.
    \item F1 score is defined as $\frac{2*(P*R)}{P+R}$.
\end{itemize}

Accuracy is measured on the entire test set, while Precision, Recall, and F1 score were measured with respect to fake news and real news respectively. 

\subsection{Experimental Setup}\label{sec:setup}
In the process of text embedding, we use pretrained BERTweet \cite{nguyen2020bertweet}, a BERT\textsubscript{base} \cite{devlin2019bert} model pretrained with RoBERTa \cite{liu2019roberta} on Twitter, for tweet text, and a T5 \cite{raffel2020exploring} model fine-tuned on news datasets for news text, \footnote{The pre-trained T5 is available at \url{https://huggingface.co/mrm8488/t5-base-finetuned-summarize-news}.} and all of which are implemented by HuggingFace \cite{wolf2019huggingface} with dimension 768.
In the procedure of image encoding, ResNet \cite{resnet} is used.
For RWR, we set $p=0.5$, $\mathcal{T}=10000$ for all datasets, $\gamma=15$ for PHEME, $\gamma=200$ for PolitiFact, and $\gamma=30$ for GossipCop.

\newcommand{\svmgamma}{\mathit{gamma}}
For HetTransformer, we set the number of attention heads as $8$ and $lr=1e-3$ for all the datasets. The number of encoder and decoder layers is set as $6$ for PolitiFact, and $1$ for PHEME and GossipCop respectively.
In alignment with the original papers, 
SVM-RBF \cite{yang2012automatic} and DTC \cite{castillo2011information} rely on human-engineered features only.
SVM-RMF achieves highest test accuracy with $\svmgamma=7e-7$, $C=1.3$ for PolitiFact, $\svmgamma=1e-4$, $C=100$ for GossipCop, and $\svmgamma=1e-7$, $C=120$ for PHEME, and 
for DTC, we use entropy as the criterion and it achieves highest test accuracy when the maximum depth is $d=11$ for PolitiFact and $d=9$ for GossipCop and PHEME.
For BERT \cite{devlin2019bert}, the concatenation of the pooled output from the pretrained BERT\textsubscript{base} and other hand-crafted features are inputted to one output layer for prediction.
In HMGNN \cite{zhu2020heterogeneous}, PolitiFact needs $k=6$ nearest neighbors, and GossipCop and PHEME needs $k=20$ ones.
In the neighbor aggregation for HGT \cite{hu2020heterogeneous} and HetGNN \cite{zhang2019heterogeneous}, the graph inputs are the same as ours in all datasets.
For HGT, the number of layers is $3$, and the number of heads is $8$ for all datasets. 
For HetGNN, we use the same input as HetTransformer, and the hyperparameters are the same as the default values. 
For IARNet, $5$ posts are sampled as neighbors through random walk for each news nodes in all datasets. Fine-tuned RoBERTa \cite{liu2019roberta} is adopted to generate word embeddings for textual content, and ResNet \cite{resnet} is adopted to encode image content. The number of attention heads is set to be $n=2$. The model is trained with $\svmgamma=1e-2$ for all datasets.
The other settings are the same as the original papers.

In all the experiments, we hold out 10\% of each dataset to be the test set and then split the remaining data into ratio 4:1 for train set and validation set.
The hyper-parameters are tuned according to the accuracy on the validation set via random search, and the results are evaluated on the test set.
We run each experiment for 5 times and report the performance of highest accuracy.
SGD optimizer is adopted.
Each model was trained for at most 40 epochs with a patience of 5.
The experiments are conducted on an Intel Core i9-10920X CPU and an ASUS RTX2080Ti Turbo 11GD6 GPU with 32GB RAM. The average runtime for HetTransformer on PolitiFact: 33.16 seconds per epoch, GossipCop: 558.06 seconds per epoch, PHEME: 83.28 seconds per epoch.
\begin{table*}[tpb!]
\caption{Comparison with existing methods}
\begin{center}
\begin{tabular}{l l c c c c c c c}
\hline
\multirow{2}{*}{Dataset} &
\multirow{2}{*}{Method} & 
\multirow{2}{*}{Accuracy} & 
\multicolumn{3}{c}{Fake News} & \multicolumn{3}{c}{Real News} \\
\cline{4-9}
& & & Precision & Recall & F1 Score & Precision & Recall & F1 Score \\
\hline
\multirow{8}{*}{PolitiFact} 
 & SVM-RBF \cite{yang2012automatic} & 0.750 & 0.719 & 0.605 & 0.657 & 0.766 & 0.845 & 0.803 \\
 & DTC \cite{castillo2011information} & 0.760 & 0.647 & 0.868 & 0.742 & 0.889 & 0.690 & 0.777 \\ 
\cline{2-9} 
 & IARNet \cite{yu2020iarnet} & 0.900 & 0.909 & \textbf{0.952} & 0.930 & 0.943 & 0.892 & 0.917\\
 & HMGNN \cite{zhu2020heterogeneous} & 0.884 & 0.868 & 0.846 & 0.857 & 0.895 & 0.911 & 0.903 \\
& HetGNN \cite{zhang2019heterogeneous} & 0.925 & 0.971 & 0.872 & 0.920 & 0.889 & 0.976 & 0.930\\
\cline{2-9}
 & BERT \cite{devlin2019bert} & 0.887 & 0.837 & 0.900 & 0.868 & 0.926 & 0.877 & 0.901 \\
 & HGT \cite{hu2020heterogeneous} & 0.900 & 0.861 & 0.949 & 0.902 & 0.946 & 0.854 & 0.897\\
 & \textbf{HetTransformer} & \textbf{0.975} & \textbf{1.000} & 0.949 & \textbf{0.974} & \textbf{0.954} & \textbf{1.000} & \textbf{0.976}\\
\hline
\multirow{8}{*}{GossipCop} 
 & SVM-RBF \cite{yang2012automatic} & 0.885 & 0.793 & 0.737 & 0.764 & 0.913 & 0.935 & 0.924 \\
 & DTC \cite{castillo2011information} & 0.882 & 0.832 & 0.665 & 0.739 & 0.894 & 0.955 & 0.924 \\
\cline{2-9}
 & IARNet \cite{yu2020iarnet} & 0.942 & 0.848 & 0.922 & 0.883 & 0.974 & 0.946 &0.960 \\
 & HMGNN \cite{zhu2020heterogeneous} & 0.939 & 0.876 & 0.876 & 0.876 & 0.960 & 0.960 & 0.960 \\
 & HetGNN \cite{zhang2019heterogeneous} & 0.962 & 0.930 & 0.913 & 0.922 & 0.972 & 0.978 & 0.975 \\
\cline{2-9}
 & BERT \cite{devlin2019bert} & 0.935 & 0.810 & 0.940 & 0.870 & 0.981 & 0.933 & 0.956 \\
 & HGT \cite{hu2020heterogeneous}& 0.960 & 0.919 & 0.915 & 0.917 & 0.973 & 0.974 & 0.973\\
 & \textbf{HetTransformer} & \textbf{0.990} & \textbf{0.978} & \textbf{0.980} & \textbf{0.979} & \textbf{0.994} & \textbf{0.993} & \textbf{0.993} \\
\hline
\multirow{8}{*}{PHEME} 
 & SVM-RBF \cite{yang2012automatic} & 0.705 & 0.631 & 0.517 & 0.568 & 0.737 & 0.818 & 0.775 \\
 & DTC \cite{castillo2011information} & 0.713 & 0.670 & 0.555 & 0.607 & 0.736 & 0.819 & 0.775 \\
\cline{2-9}
 & IARNet \cite{yu2020iarnet} & 0.811 & 0.740 & 0.718 & 0.729 & 0.841 & 0.856 & 0.848 \\
 & HMGNN \cite{zhu2020heterogeneous} & 0.767 & 0.684 & 0.696 & 0.690 & 0.817 & 0.809 & 0.813 \\
 & HetGNN \cite{zhang2019heterogeneous} & 0.817 & \textbf{0.773} & 0.722 & 0.747 & 0.840 & \textbf{0.873} & 0.856 \\
\cline{2-9}
 & BERT \cite{devlin2019bert} & 0.768 & 0.673 & 0.716 & 0.694 & 0.829 & 0.799 & 0.814 \\
 & HGT \cite{hu2020heterogeneous} & 0.773 & 0.689 & 0.718 & 0.703 & 0.827 & 0.807 & 0.817\\
 & \textbf{HetTransformer} & \textbf{0.825} & 0.756 & \textbf{0.784} & \textbf{0.770} & \textbf{0.868} & 0.849 & \textbf{0.858} \\
\hline
\end{tabular}
\label{tab:compare-existing}
\end{center}
\end{table*}
\subsection{Performance Analysis}\label{sec:performance}
Table \ref{tab:compare-existing} shows the quantitative comparison of HetTransformer with the baselines, where the best results are highlighted in bold.
HetTransformer consistently outperforms state-of-the-art baselines on all of the benchmark datasets in accuracy, F1 score for fake news, and F1 score for real news, showing that our approach is potentially effective and robust.

Specifically, HetTransformer achieves accuracies of 97.5\% on PolitiFact, 99.0\% on GossipCop and 82.5\% on PHEME, which demonstrates the strong flexibility of our model in adapting to different datasets. The excellent performance shows that the idea of combining the heterogeneous content encoder and Transformer-based heterogeneous neighbor aggregator can effectively learn the semantics of nodes in social network. 

It is obvious to notice that the performance of traditional hand-crafted methods is relatively weak (SVM-RBF \cite{yang2012automatic} and DTC \cite{castillo2011information}) in comparison to other methods that do not focus on feature engineering. This can be attribute to the fact that the traditional hand-crafted methods focus on manually extracting features from news content without capturing global structural representation of the news propagation pattern.

In terms of the GNN-based methods (IARNet \cite{yu2020iarnet}, HMGNN \cite{zhu2020heterogeneous}, and HetGNN \cite{zhang2019heterogeneous}). HetTransformer outperforms them on most of the evaluation metrics, indicating the superior ability of HetTransformer to capture the structural context of nodes in social network. 

As for the Transformer-based methods (BERT \cite{devlin2019bert} and HGT \cite{hu2020heterogeneous}). HetTransformer outperforms them on all of the evaluation metrics in three datasets. BERT only utilizes Transformer structure to capture the textual embedding from the news content, which makes it fail to utilize the useful news propagation representation in social network. HetTransformer is different from HGT in the sense that HetTransformer utilizes multi-modal feature encoders instead of the text-only feature encoders in HGT; HetTransformer employs an encoder-decoder Transformer structure which is more effective in capturing the global structural representaiton of the target news. 

Comparison with the baselines methods from Table \ref{tab:compare-existing} also shows that: 

(1) Modeling the propagation networks in fake news detection as graphs with heterogeneous edges \textit{and} heterogeneous nodes, as in HetTransformer, HetGNN \cite{zhang2019heterogeneous}, and HGT \cite{hu2020heterogeneous}, is more informative than modeling them as the ones with homogeneous nodes, as in HMGNN \cite{zhu2020heterogeneous}.

(2) The facts that HetTransformer outperforms HGT \cite{hu2020heterogeneous}, and HetGNN \cite{zhang2019heterogeneous} outperforms HMGNN \cite{zhu2020heterogeneous}, may be caused by the ability of HetTransformer and HetGNN to fuse multi-domain inputs effectively, such as the news titles, news content, and top images.

(3) The result that HetTransformer outperforms BERT \cite{devlin2019bert}, may be attributed to the fact that HetTransformer not only uses Transformer structure to capture the textual features but also uses Transformer to aggregate the heterogeneous neighbor relationship.

From (1)--(3), HetTransformer leverages Transformer in both textual feature embedding and social context aggregation of heterogeneous edges and nodes with multi-modal input, and these advantages lead to its state-of-the-art performance on all benchmark datasets.

Overall, HetTransformer outperforms the best baseline by 5.0\% in accuracy on PolitiFact, 2.8\% on GossipCop, and 0.8\% on PHEME. HetTransformer improves the result of best baseline by 4.4\% in F1 score for fake news on PolitiFact, 5.7\% on GossipCop, and 2.3\% on PHEME. And F1 score for real news on PolitiFact is improved by 4.6\%, 1.8\% on GossipCop, and 0.2\% on PHEME. This result demonstrates the superiority of HetTransformer in learning node semantics and structural representations. 

\subsection{Ablation Study}\label{sec:ablation}
To demonstrate the effectiveness of the design choice of HetTransformer, we conduct an ablation study on its input components and model substructures.
For ``$-$ Transformer decoder,'' we remove the decoder of the Transformer in Section \ref{sec: neighbors aggregator} and merely utilize the output of the encoder of the Transformer to make the prediction, where the inputs to the encoder remain the same.
For ``$-$ self-attentions $+$ BiLSTM,'' we replace the self-attentions with BiLSTM, mimicking the design choice of HetGNN \cite{zhang2019heterogeneous}.
As for ``$-$ positional embeddings,'' we remove the positional embeddings of the input nodes to the Transformer.
The self-attentions and positional embeddings are described in Section \ref{sec:Content Aggregation} and Section \ref{sec: neighbors aggregator}.
As shown in Table \ref{tab:ablation}, altering any substructures or inputs of HetTransformer leads to degeneration of its performance. Specifically, the accuracy drops 6.2\% when decoder is removed, 5\% when self-attentions are replaced with BiLSTM and 2.5\% when positional embeddings are removed for PolitiFact dataset. For GossipCop dataset, the performance in terms of accuracy has a slight decrease of 0.2\%, 0.1\%, and 0.4\% respectively. As for PHEME, the decrease in accuracy are 1.3\%, 3.8\%, and 2.9\% with respect to the three experiments.

\begin{table*}[tb!]
\caption{Results of ablation study} 
\begin{center}
\begin{tabular}{l l c c c c c c c}
\hline
\multirow{2}{*}{Dataset} &
\multirow{2}{*}{Setting} & 
\multirow{2}{*}{Accuracy} & 
\multicolumn{3}{c}{Fake News} & \multicolumn{3}{c}{Real News} \\
\cline{4-9}
& & & Precision & Recall & F1 Score & Precision & Recall & F1 Score \\
\hline
\multirow{4}{*}{PolitiFact} 
& $-$ Transformer decoder & 0.913 & 0.905 & 0.927 & 0.916 & 0.921 & 0.897 & 0.909 \\
& $-$ Self-attentions $+$ BiLSTM & 0.925 & 0.971 & 0.872 & 0.919 & 0.889 & 0.976 & 0.930 \\
& $-$ Positional embeddings & 0.950 & 0.927 & 0.974 & 0.950 & 0.974 & 0.927 & 0.950 \\
& \textbf{HetTransformer} & \textbf{0.975} & \textbf{1.000} & \textbf{0.949} & \textbf{0.974} & \textbf{0.954} & \textbf{1.000} & \textbf{0.976}\\
\hline
\multirow{4}{*}{GossipCop}
& $-$ Transformer decoder & 0.988 & 0.967 & 0.982 & 0.975 & 0.994 & 0.989 & 0.992 \\
& $-$ Self-attentions $+$ BiLSTM & 0.989 & 0.975 & 0.980 & 0.977 & 0.994 & 0.992 & \textbf{0.993} \\
& $-$ Positional embeddings & 0.986 & 0.958 & \textbf{0.986} & 0.972 & \textbf{0.996} & 0.986 & 0.991 \\
& \textbf{HetTransformer} & \textbf{0.990} & \textbf{0.978} & 0.980 & \textbf{0.979} & 0.994 & \textbf{0.993} & \textbf{0.993} \\
\hline
\multirow{4}{*}{PHEME}
& $-$ Transformer decoder & 0.812 & 0.750 & 0.747 & 0.748 & 0.849 & \textbf{0.851} & 0.850 \\
& $-$ Self-attentions $+$ BiLSTM & 0.787 & 0.728 & 0.688 & 0.707 & 0.819 & 0.846 & 0.832 \\
& $-$ Positional embeddings  & 0.796 & 0.700 & \textbf{0.842} & 0.764 & \textbf{0.885} & 0.771 & 0.824 \\
& \textbf{HetTransformer} & \textbf{0.825} & \textbf{0.756} & 0.784 & \textbf{0.770} & 0.868 & 0.849 & \textbf{0.858} \\
\hline
\end{tabular}
\label{tab:ablation}
\end{center}
\end{table*}

\begin{figure*}
     \centering
     \begin{subfigure}[b]{0.325\textwidth}
         \centering
         \includegraphics[width=\textwidth]{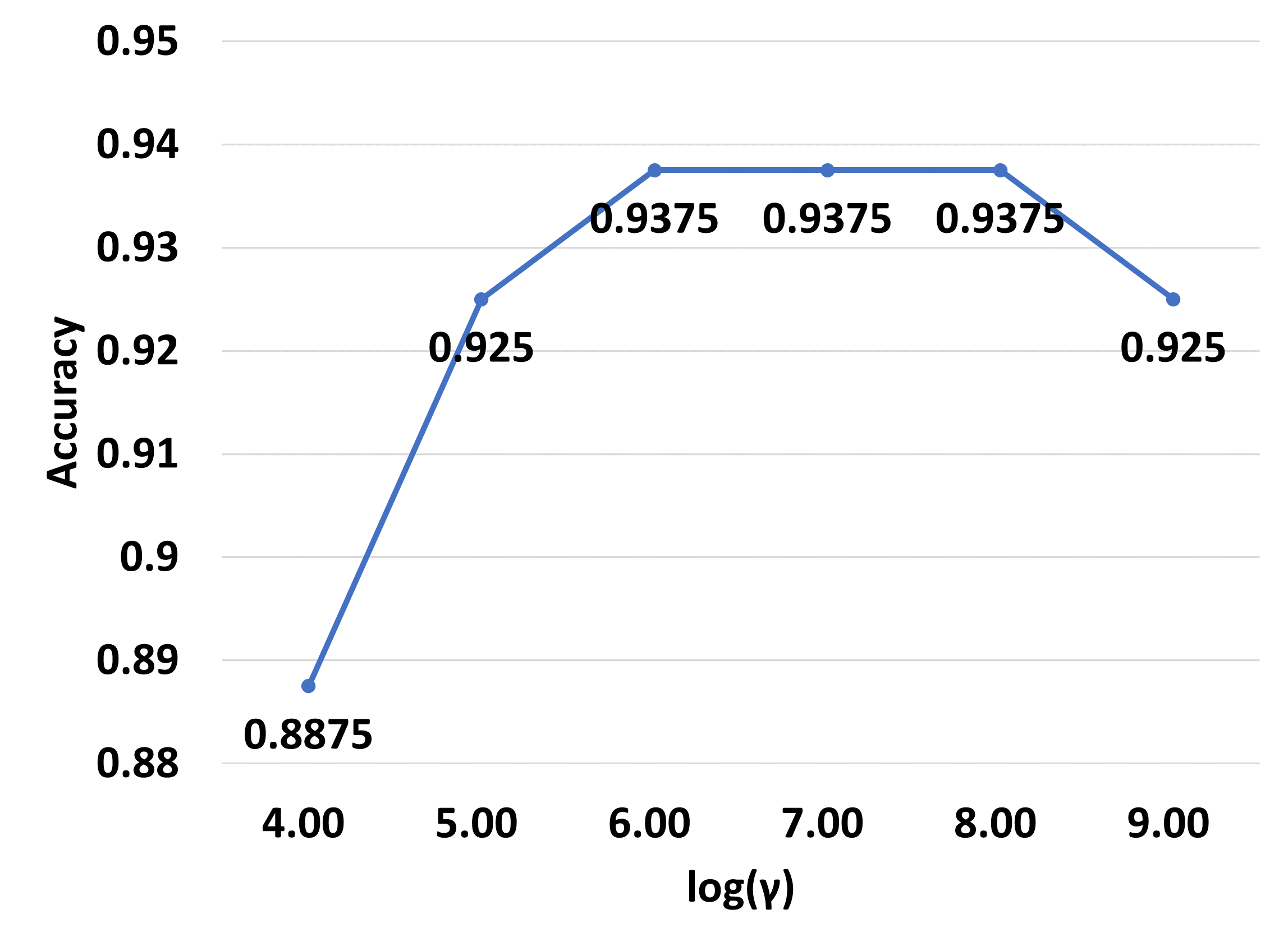}
         \caption{PolitiFact}
         \label{fig:sensitivity-politifact}
     \end{subfigure}
     \hfill
     \begin{subfigure}[b]{0.325\textwidth}
         \centering
         \includegraphics[width=\textwidth]{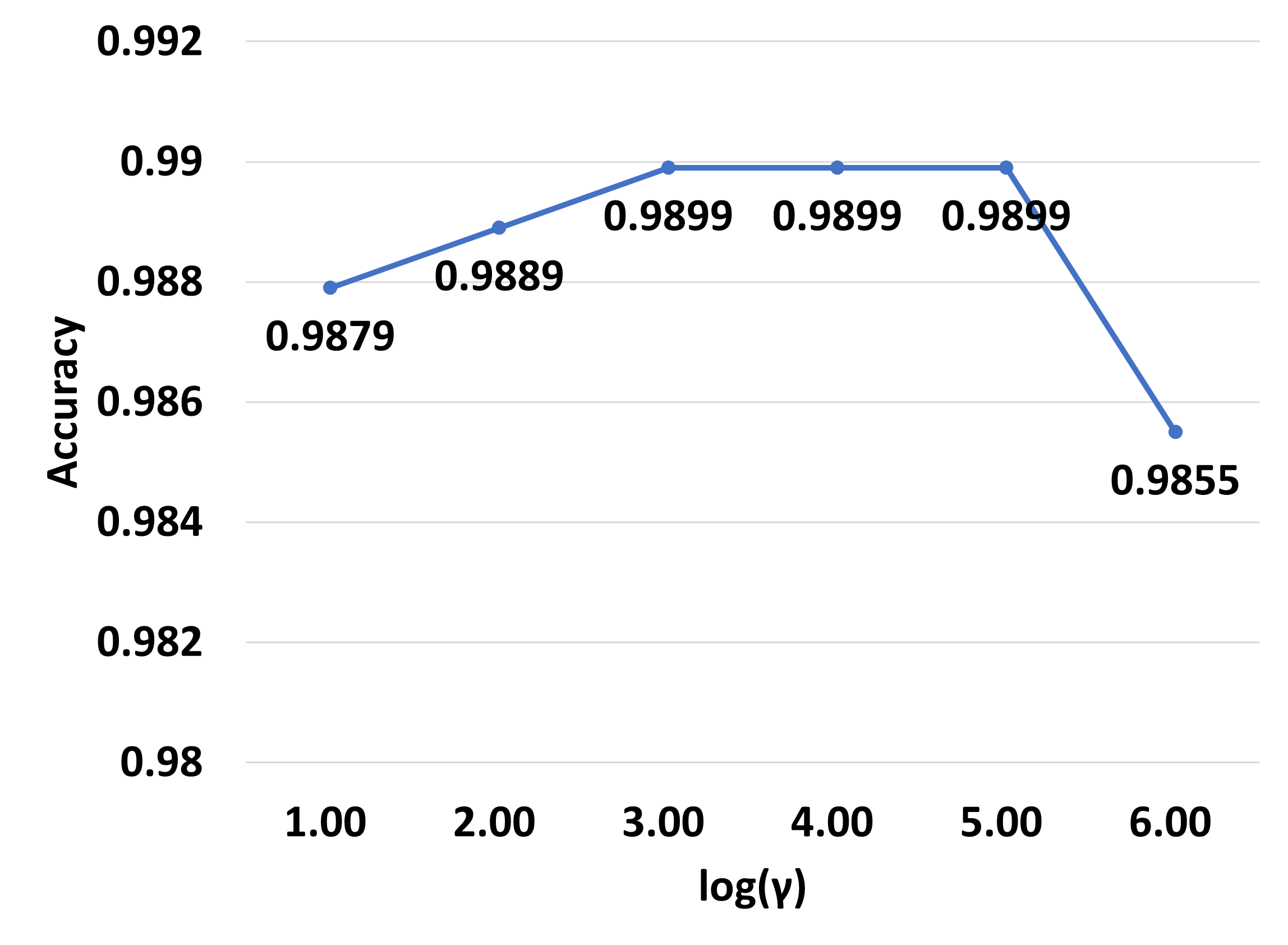}
         \caption{GossipCop}
         \label{fig:sensitivity-gossipcop}
     \end{subfigure}
     \hfill
     \begin{subfigure}[b]{0.325\textwidth}
         \centering
         \includegraphics[width=\textwidth]{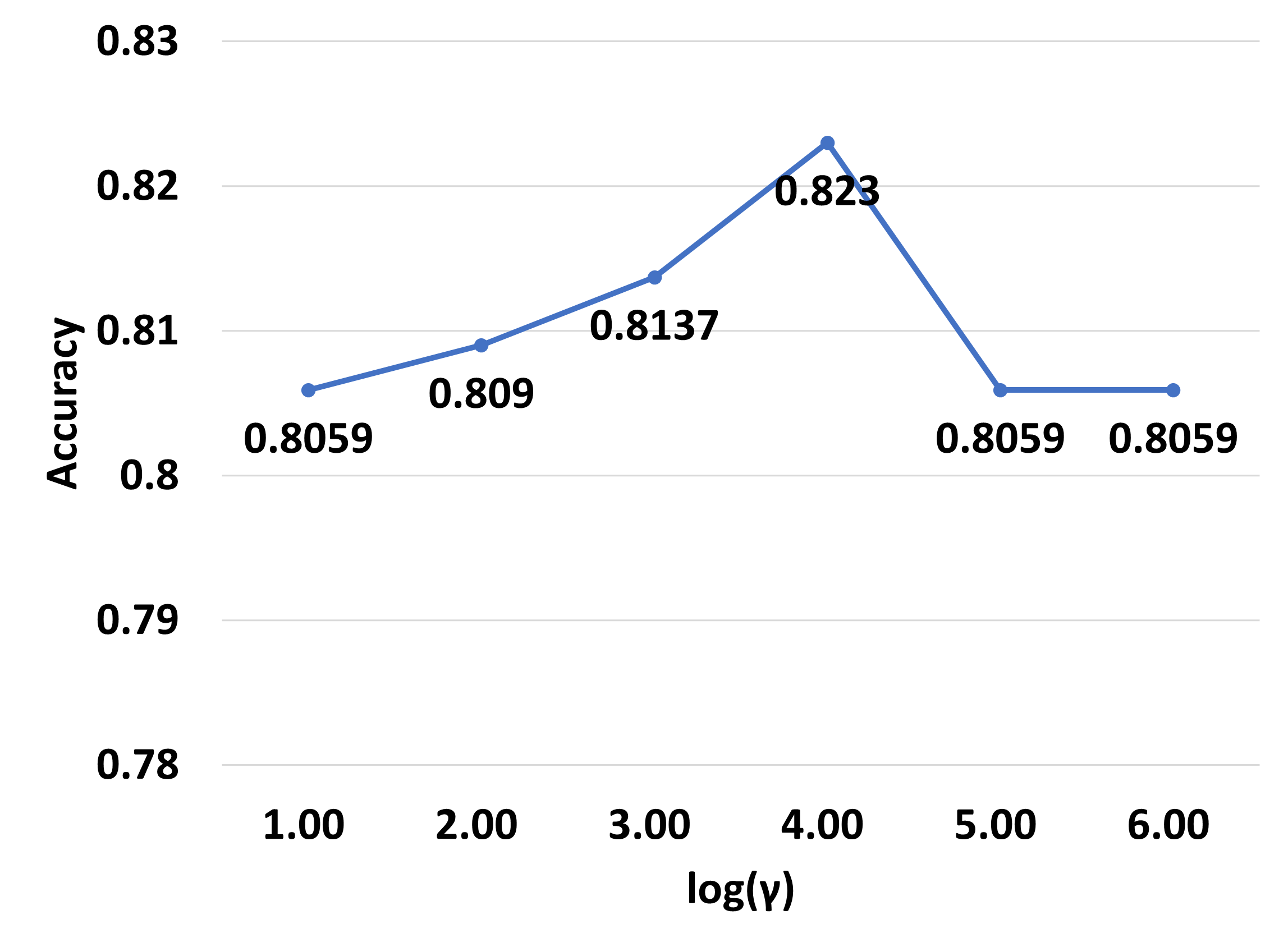}
         \caption{PHEME}
         \label{fig:sensitivity-pheme}
     \end{subfigure}
        \caption{Impact analysis of $\gamma$ in RWR for fake news detection}
        \label{fig:sensitivity-graphs}
\end{figure*}

\subsection{Parameter Sensitivity}\label{sec:sensitivity}
We experiment on the effect of $\gamma$ in RWR on PolitiFact, GossipCop, and PHEME datasets, and the result is shown in Fig. \ref{fig:sensitivity-politifact}, Fig. \ref{fig:sensitivity-gossipcop}, and Fig. \ref{fig:sensitivity-pheme}.\footnote{The $\log$ in the figures denotes binary logarithm.}
The hyperparameter $\gamma$ denotes the number of iterations in RWR, which affects the size of the neighbors sampled for each node. 
When $\gamma$ is appropriate, the sampled neighbors of the target nodes are especially representative and useful for encoding the context of the target nodes.

However, too small of a $\gamma$ limits the total amount of information available for classification and is thus harmful for the performance of HetTransformer.
A large $\gamma$, on the other hand, results in a large yet less representative neighborhood of the target node. The information brought by this kind of neighborhood may be noisy, which also hinders the performance.
Hence, the trade-off between the quality and quantity of information, can be altered with the different choices of $\gamma$.

As shown in Fig. \ref{fig:sensitivity-pheme}, initially, the accuracy of HetTransformer model increases when the value of $\gamma$ increases and then decreases as the value of $\gamma$ is too large. The highest accuracy is achieved when $\log_2(\gamma) \approx 4$ in PHEME dataset.

Similar patterns can be captured in PolitiFact dataset and GossipCop dataset as shown in Fig. \ref{fig:sensitivity-politifact} and Fig. \ref{fig:sensitivity-gossipcop}.

\section{Conclusions and Future Work}
\label{sec: conclusion}
This paper proposes HetTransformer, a novel Transformer-based model to detect fake news on heterogeneous social network graph. To the best of our knowledge, HetTransformer is the first work that applies encoder-decoder Transformer structure on fake news detection. HetTransformer utilizes multi-modal feature encoders to capture the local semantics of entities in social network, while uses encoder-decoder Transformer structure to effectively capture the global structural representation of the propagation patterns of news. Extensive experiments conducted on three real-world datasets demonstrate HetTransformer's excellent performance of detecting fake news in social network. In the future work, we plan to extend HetTransformer to other scenarios such as dynamic graph and early detection of fake news. Other settings of HetTransformer such as Semi-supervised and Active Learning versions will also be explored.
\section*{Acknowledgment}
The authors would like to thank all the anonymous reviewers for their valuable feedback.
\bibliographystyle{IEEEtranS}
\bibliography{IEEEabrv,references.bib}



\end{document}